\documentclass[]{aa501}
\usepackage{times}
\usepackage[]{graphics}
\usepackage[]{psfig}
\usepackage{graphicx}

\begin{document}

\newcommand{\kms}{km~s$^{-1}$}
\newcommand{\cm}{cm$^{-2}$}
\newcommand{\lya}{Lyman~$\alpha$}
\newcommand{\lyb}{Lyman~$\beta$}
\newcommand{\za}{$z_{\rm abs}$}
\newcommand{\ze}{$z_{\rm em}$}
\newcommand{\nhi}{$N$\/(H~I)}

\def\ltsima{$\; \buildrel < \over \sim \;$}
\def\simlt{\lower.5ex\hbox{\ltsima}}
\def\gtsima{$\; \buildrel > \over \sim \;$}
\def\simgt{\lower.5ex\hbox{\gtsima}}
\def\arcs{$''~$}
\def\arcm{$'~$}

%   \thesaurus{4(12.03.3;11.01.1;11.05.2;11.17.1)} 
%
\title{The Abundances of Nitrogen and Oxygen in \\
Damped \lya\ Systems}

\titlerunning{Nitrogen and Oxygen in Damped \lya\ Systems}

\author{Max Pettini\inst{1}
       \and Sara L. Ellison\inst{2}
       \and Jacqueline Bergeron\inst{3,4}
       \and Patrick Petitjean\inst{4,5}}

\institute{Institute of Astronomy, Madingley Road, Cambridge, CB3 0HA, UK
           \and European Southern Observatory, Casilla 19001, Santiago 19, Chile
           \and European Southern Observatory, Karl-Schwarzschild-Strasse 2, 
           Garching bei M\"{u}nchen, D-85748, Germany
           \and Institut d'Astrophysique de Paris, 98bis Boulevard d'Arago, F- 75014 Paris, France 
           \and LERMA, Observatoire de Paris, 61 Avenue de l'Observatoire, F-75014 Paris, France
}

\date{Received / Accepted}

\abstract{
We take a fresh look at the abundance of 
nitrogen in damped \lya\ systems (DLAs) 
with oxygen abundances between 
$\sim$1/10 and $\sim$1/100 of solar. 
This is a metallicity regime poorly sampled
in the local universe and where QSO absorbers
may hold clues to both
the nucleosynthetic origin of N and the chemical 
evolution of high redshift galaxies.
We combine new VLT UVES observations %of three QSOs
with others from the literature to form a sample
of 10 DLAs in which the abundances of N 
and of one of the two $\alpha$-capture elements
O or S have been measured. The sample consists
exclusively of high resolution, high signal-to-noise ratio
data obtained with 8--10\,m class telescopes.
We confirm earlier reports that the (N/O) ratio
exhibits a larger range of values than other ratios
of heavy elements in DLAs; however, all 10 DLAs
fall in the region of the (N/O) vs. (O/H)
plot delimited by the primary and secondary levels
of nitrogen production.
Our results provide empirical evidence
in support of the proposal that intermediate mass stars
($4 \simlt M/M_{\odot} \simlt 7$)
are the main source of primary nitrogen,
so that its release 
into the interstellar medium lags
behind that of oxygen, which is
produced by Type II supernovae.
A high proportion (40\%) of the DLAs in our sample 
have apparently not yet attained the full 
primary level of N enrichment; this finding may 
be an indication that the metallicity
regime we are studying preferentially
picks out galaxies which have only recently
condensed out of the intergalactic medium and begun
forming stars. Alternatively, the delay in the 
release of N following an episode of star formation may
increase with decreasing metallicity, if stars
of progressively lower masses than $4\,M_{\odot}$ 
can synthesize N in their hotter interiors.
In this general picture, the uniform value of (N/O) 
measured locally in metal-poor star-forming
galaxies implies that they are not young,
and is consistent with the presence of older 
stellar populations revealed by imaging
studies with the {\it Hubble Space Telescope}.
\vspace{-0.1cm}
}

\maketitle

\section{Introduction}

The nucleosynthetic origin of nitrogen continues to be a subject
of considerable interest and discussion. There is general
agreement that the main pathway is a six step process in the CN
branch of the CNO cycle which takes place in the stellar H
burning layer, with the net result that $^{14}$N is synthesised
from $^{12}$C and $^{16}$O. The continuing debate, however,
centres on which range of stellar masses is responsible for the
bulk of the nitrogen production.
A comprehensive reappraisal of the problem
was presented by Henry, Edmunds, \& K\"{o}ppen
(2000) who compiled an extensive set of abundance
measurements and computed chemical evolution models using
published yields.  Briefly, nitrogen has both a primary and a
secondary component, depending on whether the seed carbon and
oxygen are those manufactured by the star during helium burning,
or were already present when the star first condensed out of the 
interstellar medium (ISM). 

%
% TABLE 1
%___________________________________ Two column table (place early!)
 
\begin{table*}[!]
\begin{center}
\caption{Damped Ly$\alpha$ Systems Observed} 
\begin{tabular}{lcccccccc}
\hline\hline
\noalign{\smallskip}
QSO
& V
& $z_{\rm em}$
& $z_{\rm abs}$
& $N$\/(H~I)$^{a}$
& Exp. Time
& Resolution
& Wavelength Range
& S/N$^{b}$\\

& (mag)
& 
&  
& (cm$^{-2}$)
& (s)
& (km~s$^{-1}$)
& (\AA) 
&
\smallskip
\\
\hline
\\
Q1409$+$095   & 18.6    & 2.856  & 2.45620  & $3.5  \times 10^{20}$ & 16\,200 & 7 & 3793--4989, 6726--10\,255$^{c}$ & 15--25\\
%Q1409$+$095   & 18.6    & 2.856  & 2.66820  & $5.0  \times 10^{19}$ & 16\,200 & 7 & 3793--4989, 6726--10\,255$^{c}$ & 15--25\\
Q1444$+$014   & 18.5    & 2.206  & 2.08681  & $1.6  \times 10^{20}$ & 18\,000 & 7 & 3295--6650$^{d}$                & 10--20\\
Q2206$-$199   & 17.33   & 2.559  & 2.07623  & $2.7  \times 10^{20}$  & 24\,300 & 7 & 3295--10\,255$^{d}$            & 20--25\\ 
\\
\hline
\end{tabular}
\begin{minipage}{160mm}
\smallskip
\item[$^{\rm a}$] Determined by profile fitting to the damping wings 
of the \lya\ absorption line (see section 2 and Figure 1). Typical
error in $N$(H~I) is $\pm 10\%$.
\item[$^{\rm b}$] The signal-to-noise ratio varies along each spectrum. These values
(per pixel) refer to the QSO continuum in the regions of the N~I~$\lambda\lambda 1200$ and O~I~$\lambda 1302.2$ lines.
\item[$^{\rm c}$] With a wavelength gap between 8520 and 8667 \AA.
\item[$^{\rm d}$] With some wavelength gaps.
\end{minipage}
\end{center}
\end{table*}

% ------------------------------------------------------------------

\subsection{Nitrogen and Oxygen in H~II regions} 

Observational evidence for this dual nature of nitrogen
is provided mainly from measurements of the N and O abundances
in H~II regions. 
(For simplicity in this paper we use parentheses
to indicate logarithmic ratios of number densities; 
adopting the recent reappraisal of solar photospheric
abundances by Holweger (2001), we have (N/H)$_{\odot} = -4.07$;
(O/H)$_{\odot} = -3.26$; and (N/O)$_{\odot} = -0.81$).
In H~II regions of nearby galaxies, (N/O)
exhibits a strong dependence on (O/H) when the latter is greater
than $\sim 2/5$ solar; this is generally interpreted 
as the regime where secondary N becomes 
dominant.\footnote{As an aside, Henry et al. (2000) 
pointed out that the rise
in (N/O) with (O/H) is steeper than would be normally expected
for a purely secondary element and proposed that the effect is augmented
by a decreasing O yield with increasing metallicity (Maeder
1992).} At low metallicities on the other hand, 
when (O/H)$\,\simlt -4.0$ (that is, $\simlt 1/5$ solar), 
N is mostly primary and tracks O; this results in a %flat
plateau at (N/O)~$\simeq -1.5$. 

The principal sources of primary N are thought to be intermediate
mass stars ($4 \simlt M/M_{\odot} \simlt 7$) during the 
asymptotic giant branch (AGB) phase. 
Henry et al. (2000) showed that, integrating
along the IMF the N yields from intermediate mass stars by van
den Hoek \& Groenewegen (1997) and the O yields from massive
stars by Maeder (1992) at metallicity $Z = 1/20 Z_{\odot}$, one
obtains (N/O)\,$ = -1.41$ in good agreement with the 
observed plateau.

A corollary of the hypothesis that intermediate mass
stars are the main producers of primary nitrogen is that
its release in the ISM should lag behind that of O,
since the latter is widely believed to be
produced by massive stars which explode 
as Type II supernovae (SN) 
soon after an episode of star formation.
Henry et al. (2000) calculated this time delay 
to be approximately 250\,Myr; at low
metallicities the (N/O) ratio could then perhaps be used as a
clock with which to measure the past rate of star formation,
as proposed by Edmunds \& Pagel (1978).
Specifically, in metal-poor galaxies which have 
only recently experienced a 
burst of star formation one may expect to find
values of (N/O) {\it below} the primary plateau
at (N/O)~$\simeq -1.5$.

This scenario has been discussed at length over the 
last few years with claims of both conflicting and 
supporting evidence from the data. 
Izotov \& Thuan (1999) and Izotov et al. (2001) were struck by the
constant values of all the element ratios, including (N/O), which they
measured in blue compact dwarf (BCD) galaxies with (O/H)\,$ \simlt -4.4$ 
(corresponding to an oxygen abundance less than $\sim$1/15 of solar).
This finding led them to the radical
proposal that (a) these are galaxies undergoing their first burst
of star formation, and (b) all the elements they observed have a
primary origin in massive stars, so that they are released at the
same time as O, thereby disposing altogether of the notion of a
time delay. On the other hand, a variety of studies
using high resolution {\it Hubble Space Telescope} ({\it HST})
images have shown that many of these BCDs host old
stellar populations, with ages greater than 1\,Gyr
(e.g. Schulte-Ladbeck et al. 2001 and references therein; 
Crone et al. 2002). Generally, these low mass galaxies
tend to have low rates of star formation
(even though they are observed during a bursting phase).
If the lag  between O and N production is only 
250~Myr, as proposed by Henry et al. (2000), it is
perhaps not surprising that
most of them lie near the primary value of the (N/O)
ratio expected from intermediate mass stars
(Pilyugin 1999).

In contrast, a recent survey by Contini et al. (2002)
found that UV-selected galaxies at intermediate redshifts
(0$<${\it z}$<$0.4) exhibit a wide range of (N/O) values
at a given (O/H), in many cases well below the 
primary level. While Contini et al. interpret
their results as evidence in favour of a delayed
production of primary N, one may well 
wonder at the high fraction of galaxies
apparently caught within 250\,Myr since the
last major episode of oxygen enrichment. Possibly
the UV-continuum selection technique strongly
favours galaxies with high rates of star formation
in the recent past.

\subsection{Nitrogen and Oxygen in QSO Absorption Line Systems} 

As pointed out by Pettini, Lipman, \& Hunstead (1995), 
clues to the nucleosynthetic origin 
of nitrogen can also be provided by
QSO absorption line systems, particularly the high column density 
($N$\/(H~I)~$\geq 2 \times 10^{20}~$cm$^{-2}$)
damped \lya\ absorbers (DLAs). These are
thought to represent an early stage in the evolution of galaxies,
when most of their baryonic mass was in the interstellar medium.
Apart from the obvious interest in measuring
element abundances in the distant past, when galaxies
were young, one of the advantages of DLAs is that they are generally of low
metallicity, approximately between $1/10$ and $1/100$ solar
(Pettini et al. 1999; Prochaska \& Wolfe 2002). Thus, they 
probe a regime where local H~II region abundance measurements are
sparse or non-existent and where the effect of a delayed
production of primary nitrogen should be most pronounced.

There are practical difficulties, however, in measuring
(N/O) in DLAs. The resonance lines most easily accessible,
the N~I triplet $\lambda\lambda 1199.5, 1200.2, 1200.7$
and O~I $\lambda 1302.2$, have very different optical depths.
While the latter is normally saturated in DLAs, the former
is weak and can be difficult to detect, particularly
as it falls within the \lya\ forest.
Pettini et al. (1995), and subsequently Lu, Sargent, \& Barlow
(1998) and Centuri\'{o}n et al. (1998), attempted to circumvent
the first problem by using the abundances of other $\alpha$-capture
elements, such as Si and S, as proxies for (O/H). These
studies found that the (N/$\alpha$) ratio in DLAs spans a wide
range of values, broadly within the range
bracketed by the predictions 
for primary and secondary production of nitrogen.
However, these results have been criticized by
Izotov \& Thuan (1999) and more recently
Izotov, Schaerer, \& Charbonnel (2001).
Unlike O and N, which are
mostly neutral in H~I gas, Si and S are singly
ionised. Since in principle Si~II and S~II can
also occur in H~II regions, Izotov and collaborators
have proposed that unaccounted ionisation corrections,
rather than real abundance variations, are the cause of
the observed scatter of the (N/$\alpha$) ratios
in DLAs. Their model, however, makes the radical assumption
that essentially all the metals are located in
(presumably self-enriched) H~II regions, and that
the metallicity of the neutral phase is negligible.

\subsection{Motivations for the Present Study}

In an attempt to clarify this confusing state of affairs,
we have begun a new programme of observations with
the Ultraviolet-Visual Echelle Spectrograph (UVES)
on the VLT (Kueyen) telescope (Dekker et al. 2000).
In order to minimize the practical difficulties 
outlined above, we targeted DLAs which (a) have
relatively low values of hydrogen column density
($N$\/(H~I)$\simlt 4 \times 10^{20}$\,cm$^{-2}$),
thus increasing the probability that 
the O~I~$\lambda 1302.2$ line may not be
strongly saturated), and (b) are at relatively low
redshifts ($z_{\rm abs} \simlt 2.7$), 
where the \lya\ forest begins to
thin out. Despite this observing strategy, which
capitalises on the superior performance of UVES
at ultraviolet wavelengths (D'Odorico et al. 2000),
our endeavours have been only partially successful,
as we shall see. In addition, we bring together
a number of relevant abundance measurements
which have been published since the surveys
by Lu et al. (1998) and Centuri\'{o}n et al. (1998)
for a comprehensive reappraisal of the
abundances of oxygen and nitrogen in DLAs.

\section{Observations and Data Reduction}
Details of the DLAs observed are collected in Table 1.
The data were secured over the three nights of 28--30 May 2000.
By using different cross-disperser gratings and dichroic
filters we recorded the spectra of three QSOs 
%(one with two DLAs)
over the wavelength ranges listed
in the penultimate column of Table 1. With a 1 arcsec wide
entrance slit and $2 \times 2$ binning on the CCDs,
the resulting spectral resolution was between
6.7 and 7.1\,km~s$^{-1}$ FWHM, measured from
the widths of emission lines of the Th-Ar
hollow-cathode lamp used for wavelength calibration.
The observations consisted of a number of one hour long
exposures; the QSOs were moved along the slit
between exposures so as to use different portions
of the detectors to record the spectrum of each object.
The slit was aligned at the parallactic angle throughout
the observations.

The two-dimensional images were processed with
the UVES standard pipeline software; see for example
Ellison, Ryan, \& Prochaska (2001) for a description
of the main steps.  The individual spectra extracted for each QSO
were mapped onto a common, linear, vacuum heliocentric,
wavelength scale; the interval between successive wavelength bins was set 
to the average value of the original CCD pixels, 
typically $\delta \lambda = 0.03\,$\AA\
and 0.04\,\AA\ in the blue and red wavelength regions respectively.
The individual extractions were then co-added,
with weights proportional to the S/N ratios,
to produce final blue and red spectra
for each QSO. With several independent exposures at our disposal,
we were able to exclude from the sum wavelength bins which 
deviated from the mean by several standard deviations,
because of residuals from the subtraction of cosmic-ray events, 
strong sky emission lines, and other invalid pixels.

Figure 1 shows portions of the final spectra 
in the regions near the damped \lya\ lines at $z_{\rm abs}=2.45620$ 
in Q1409$+$095, $z_{\rm abs}=2.08681$ in Q1444$+$014, 
and $z_{\rm abs}=2.07623$ in Q2206$-$199, after division by the
underlying QSO continuum. Superposed on the data are
theoretical damped absorption profiles corresponding to 
the values of $N$\/(H~I) listed in Table 1; the typical 
error in the determination of $N$\/(H~I) is about 10\%.
Our observations revealed the existence of a second
high column density absorption system in Q1409$+$095,
at $z_{\rm abs} = 2.66820$. For completeness, we have included
it in Figure 1; the profile fit shown is for
$N$(H~I)\,$= 5 \times 10^{19}$\,cm$^{-2}$.

% FIGURE 1

\begin{figure*}[h]
\centerline{\resizebox{18cm}{!}{\includegraphics{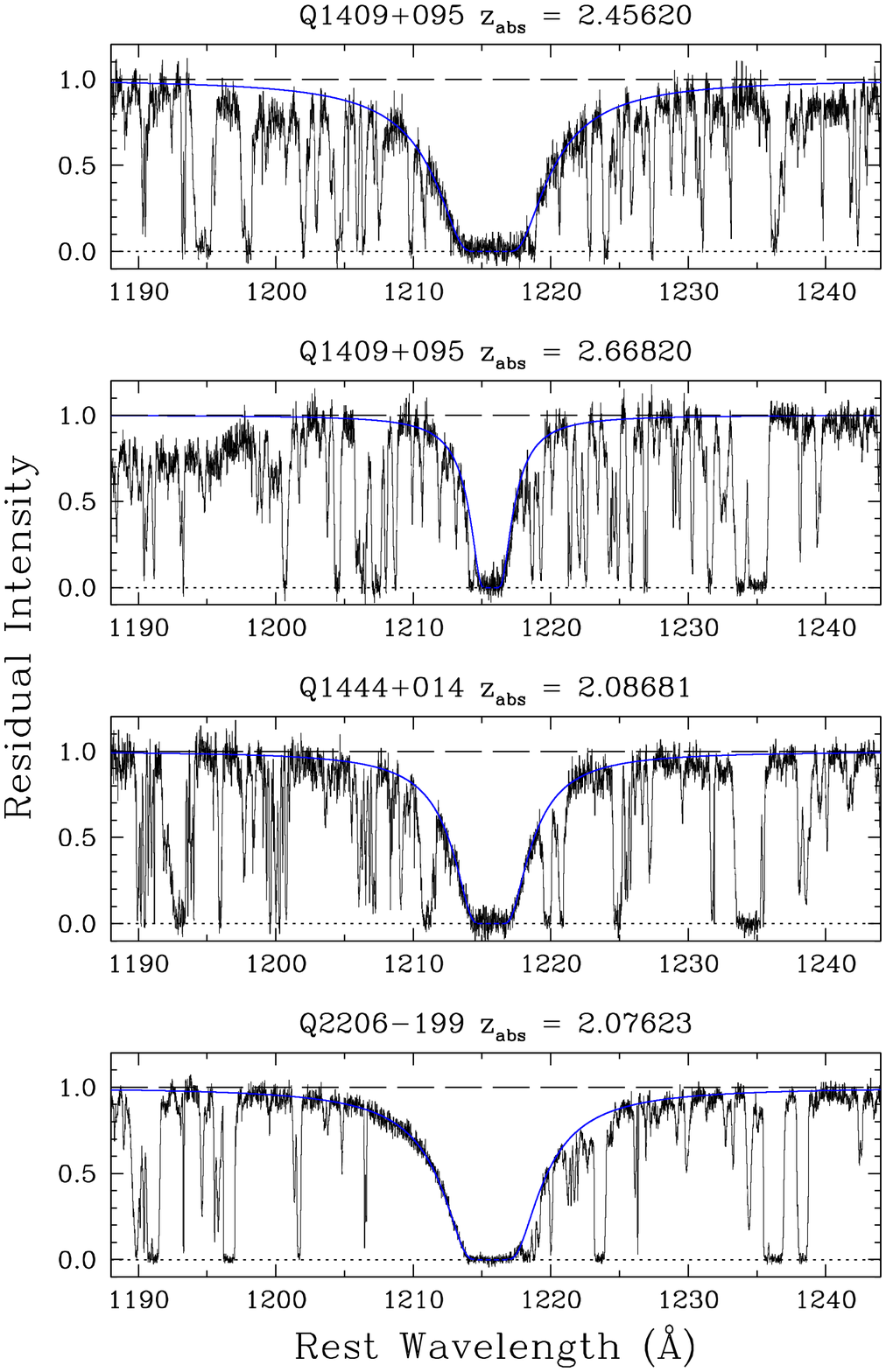}}}
\caption{Normalised UVES spectra of the three QSOs observed 
showing the regions
of the four damped \lya\ lines. Superposed on the data are
theoretical absorption profiles for \lya\ lines with the
neutral hydrogen column densities $N$\/(H~I) listed in Table 1.
}
\end{figure*}

\section{Metal Absorption Lines in the Four DLAs}

\subsection{Rationale}

As is the case with modern echelle spectra, our
observations of the three QSOs in Table 1 provide
a wealth of information on several absorption
systems, as well as on the \lya\ forest.
In the present study we are primarily interested
in the abundances of nitrogen and oxygen, detected respectively
via the N~I~$\lambda \lambda 1199.5, 1200.2, 1200.7$
triplet and the O~I~$\lambda 1302.2$ line.
Also of relevance to our analysis are the abundances of
silicon (measured via Si~II~$\lambda 1304.4$) 
and iron (several absorption lines of Fe~II throughout the 
spectral regions covered). 
Silicon is an $\alpha$-capture element
which can provide an additional check 
on the oxygen abundance in cases where
the O~I~$\lambda 1302$ line is saturated
and the corresponding value of $N$\/(O~I) is
uncertain. Sulphur would also be useful
in this context, but unfortunately the relevant absorption lines
are either blended or undetected in most of the cases
considered here. Iron is of interest because it is thought 
to be produced primarily by Type Ia supernovae;
like nitrogen, its release into the interstellar medium
should be delayed relative to that of oxygen.
The four elements in question, N, O, Si, and Fe, respond
differently to depletion onto dust grains.
In the local ISM, N and O are normally present in the gas phase
with near-solar abundances; Si and Fe, on the other hand,
are susceptible to dust depletion, the latter normally
more than the former (e.g. Savage \& Sembach 1996). Thus, by
considering all four elements together, it may be possible
to decouple nucleosynthesis and depletion effects.

\subsection{Profile Fitting and Abundance Determinations}

%
% TABLE 2
%___________________________________ Two column table (place early!)

\begin{table*}[!]
\begin{center}
\caption{Component Structure of the Metal Absorption Lines} 
\begin{tabular}{lccccccc}
\hline\hline
\noalign{\smallskip}
QSO
& Component
& $z_{\rm abs}$
& $b$
& log $N$\/(O~I)
& log $N$\/(Si~II)
& log $N$\/(Fe~II)
& log $N$\/(N~I)  \\
  
& Number
& 
& (km~s$^{-1}$)
& (cm$^{-2}$)
& (cm$^{-2}$)
& (cm$^{-2}$)
& (cm$^{-2}$) 
\smallskip
\\
\hline
\\
Q1409$+$095 & 1 & 2.45597 & 12.9 & $14.76 \pm 0.1$  & $13.82 \pm 0.03$ & $13.46 \pm 0.03$ & $\leq 12.80$ \\
            & 2 & 2.45630 & 5.8  & $14.27 \pm 0.03$ & $13.08 \pm 0.03$ & $12.81 \pm 0.03$ & $\leq 12.50$ \\
            & 3 & 2.45648 & 7.3  & $14.82 \pm 0.1$  & $13.63 \pm 0.03$ & $13.29 \pm 0.03$ & $\leq 12.77$ \\
\\
Q1409$+$095 & 1 & 2.66820 & 7.5  & $15.27 \pm 0.1$  & $14.02 \pm 0.03$ & $13.85 \pm 0.03$ & $\leq 13.45$ \\
\\
Q2206$-$199 & 1 & 2.07623 & 6.1  & $15.24 \pm 0.09$ & $13.65 \pm 0.02$ & $13.37 \pm 0.02$ & $\leq 12.88$ \\
\\
\hline
\end{tabular}
\end{center}
\end{table*}

%-------------------------------------------  

Normalised profiles of the absorption lines mentioned
above were fitted with 
Voigt profiles using the VPFIT package.\footnote{VPFIT
is available at http://www.ast.cam.ac.uk/\~\,rfc/vpfit.html}
VPFIT deconvolves the composite absorption profiles 
into the minimum number of discrete components 
and returns for each the most likely values of 
redshift $z$, Doppler width $b$ (km~s$^{-1}$),
and column density $N$ (cm$^{-2}$)
by minimizing the difference 
between observed and computed profiles.
The profile decomposition takes into
account the instrumental
point spread function of UVES.
As we shall see shortly, the velocity structure
of three out of the four DLAs studied here is very simple.
In each case the same set of values of $z$ and $b$ 
was found to provide a satisfactory fit to the absorption
lines of O~I, Si~II, and Fe~II.
Oscillator strengths and rest wavelengths were adopted from
the up to date compilation of atomic parameters maintained
by J. X. Prochaska at 
http://kingpin.ucsd.edu/\~\,hiresdla/atomic.dat
(see also Prochaska et al. 2001).
The solutions returned by VPFIT are collected in Table 2.

The error estimates $\sigma$(log\,$N$) returned by VPFIT
were typically
$< 0.05$\,dex for the optically thin Si~II and Fe~II lines
and $< 0.1$\,dex for the saturated O~I lines.
To test whether these errors are realistic 
we performed Monte Carlo type simulations similar to those
described by Bowen, Blades, \& Pettini
(1995) and Bowen, Pettini, \& Blades (2002).
Briefly, for each DLA
we used as the starting point the theoretical spectrum
produced by VPFIT with the best fit parameters
given in Table 2. From this template we generated 300
spectra with the same S/N as the original data;
these were then refitted to deduce
values of $b$ and $N$. 
From gaussian fits to the distributions
of 300 values of $b$ and $N$ we determined
values of $\sigma$($b$) and $\sigma$(log\,$N$).
The former are typically $\pm 0.2$\,\kms;
the latter, which are comparable to the values
returned by VPFIT, are listed in Table 2.
Despite the good agreement between the two methods, 
it must be remembered that these errors reflect primarily
the S/N of the data and do not include the possibility
that the velocity dispersion parameter $b$ may be
different between the different ions considered.
The consequences of letting $b$ vary from ion to ion
are discussed below for each DLA individually.

The species observed are the major ionisation stages
of their respective elements in H~I regions. Thus, to
obtain their abundances relative to hydrogen we simply
divide the column densities
returned by VPFIT by $N$\/(H~I) 
(after adding together the values for
individual velocity components within the same
DLA, when more than one component is present).
Finally these values are referred to the solar
abundance scale (Holweger 2001). The results are collected
in Table 3, where the solar reference values are 
listed in the footnotes.

This procedure would overestimate the abundances
of Si and Fe if some of the Si~II and Fe~II absorption
we detect were due to ionised gas associated
with the neutral DLA. Conversely, the abundances
of O and N would be underestimated if these
elements were partly ionised in the H~I region.
In the cases under consideration, however,
there is little evidence that these reservations
are justified. 
First, all four species exhibit very similar velocity
structure; we see no indication of components
present in the first ions and missing in the neutrals,
or vice versa. Second, the recent reappraisal
of this problem by Vladilo et al. (2001)
shows that at the values of $N$\/(H~I) of the
DLAs studied here, the ionisation corrections
to the abundances determined as above are smaller
than the typical error of $\pm 0.15$\, dex of our
abundance estimates. 
%One exception may be the
%$z_{\rm abs} = 2.66820$ DLA in Q1409$+$095
%for which the neutral hydrogen column density
%is only $N$\/(H~I)$ = 5.0 \times 10^{19}$\,cm$^{-2}$,
%but even here we see no evidence for abundance
%anomalies between the neutrals and the first ions,
%as discussed below.
We now briefly comment on each DLA in turn.

%
% TABLE 3
%___________________________________ Two column table (place early!)

\begin{table*}[!]
\begin{center}
\caption{Ion Column Densities and Element Abundances}
\begin{tabular}{lcccccccccc}
\hline\hline
\noalign{\smallskip}
QSO
& $z_{\rm abs}$
& log $N$\/(H~I)$^{a}$
& log $N$\/(O~I)
& log $N$\/(Si~II)
& log $N$\/(Fe~II)
& log $N$\/(N~I)
& [O/H]$^{b}$
& [Si/H]$^{b}$
& [Fe/H]$^{b}$
& [N/H]$^{b}$\\
  
& 
& (cm$^{-2}$)
& (cm$^{-2}$)
& (cm$^{-2}$)
& (cm$^{-2}$)
& (cm$^{-2}$)
& 
& 
& 
& 
\smallskip
\\
\hline
\\
Q1409$+$095 & 2.45620 & 20.54 & $15.15 \pm 0.06$ & $14.08 \pm 0.02$ & $13.74 \pm 0.02$ & $\leq 13.19$ & $-2.13$ & $-2.00$ & $-2.25$ & $\leq -3.28$ \\
\\
Q1409$+$095 & 2.66820 & 19.70 & $15.27 \pm 0.1$  & $14.02 \pm 0.03$ & $13.85 \pm 0.03$ & $\leq 13.45$ & $-1.17$ & $-1.22$ & $-1.30$ & $\leq -2.18$\\
\\
Q2206$-$199 & 2.07623 & 20.44 & $15.24 \pm 0.09$ & $13.65 \pm 0.02$ & $13.37 \pm 0.02$ & $\leq 12.88$ & $-1.94$ & $-2.33$ & $-2.52$ & $\leq -3.49$ \\
\\
\hline
\end{tabular}
\begin{minipage}{160mm}
\smallskip
\item[$^{\rm a}$] Typical error in log $N$(H~I) is $\pm 0.04$.
\item[$^{\rm b}$] In the usual notation: [X/H] = log(X/H) $-$ log(X/H)$_{\odot}$.
Solar abundances are from the recent updates by Holweger (2001): log(O/H)$_{\odot} = -3.264$; 
log(Si/H)$_{\odot} = -4.464$;
log(Fe/H)$_{\odot} = -4.552$; and
log(N/H)$_{\odot} = -4.069$\,.
\end{minipage}
\end{center}
\end{table*}

% ------------------------------------------------------------------

\subsection{The $z_{\rm abs} = 2.45620$ DLA in Q1409$+$095}

The Si~II~$\lambda 1304.4$ and Fe~II~$\lambda 2344.2$ 
lines in this DLA are unsaturated and are well 
reproduced with three absorption components spanning 
a velocity range $\Delta v = 44$\,\kms\ (see Figure 2
and Table 2). This is a very metal-poor system. Adding
together the column densities for the three components
we deduce that the abundance of Si is 1/100 of solar,
and that of Fe only little more than 1/200 of solar
(see Table 3). 

Even at these low abundances,
and with a hydrogen column density of 
$3.5 \times 10^{20}$\,\cm\ (near the lower limit
of $2 \times 10^{20}$\,\cm\ commonly adopted as the 
definition of a DLA), the O~I~$\lambda 1302.2$ line
is saturated (see Figure 2). If we adopt the same
set of absorption parameters as for Si~II and Fe~II,
we deduce an oxygen abundance of 1/135 of solar, in reasonable
agreement with that of Si. The good model fit
to  O~I~$\lambda 1302.2$ makes it unlikely that we
have overestimated $N$\/(O~I) (the line would be
broader than observed). The oxygen abundance could %in principle
be higher if our model underestimates the saturation
of the O~I~$\lambda 1302.2$ line; this possibility, however,
does not alter the interpretation of our results below
because it would tend to emphasize the `anomalously' low (N/O)
ratio (see \S 5).

The N~I~$\lambda 1200$ triplet at $z_{\rm abs} = 2.45620$
falls within the \lya\ forest of this $z_{\rm em} = 2.856$
QSO. Figure 2 shows the best fit to this spectral region
obtained with the model parameters of the other three lines,
but it is difficult to be confident that the weak absorption we
see, particularly in the $\lambda 1200.2$ transition, is actually
N~I and not a \lya\ forest line. Consequently, we prefer to
adopt the corresponding $N$\/(N~I) as an upper limit
to the column density of N~I, from which we deduce that
the abundance of N is less than 1/2000 of solar.

% FIGURE 2

\begin{figure*}[h]
\centerline{\resizebox{18cm}{!}{\includegraphics{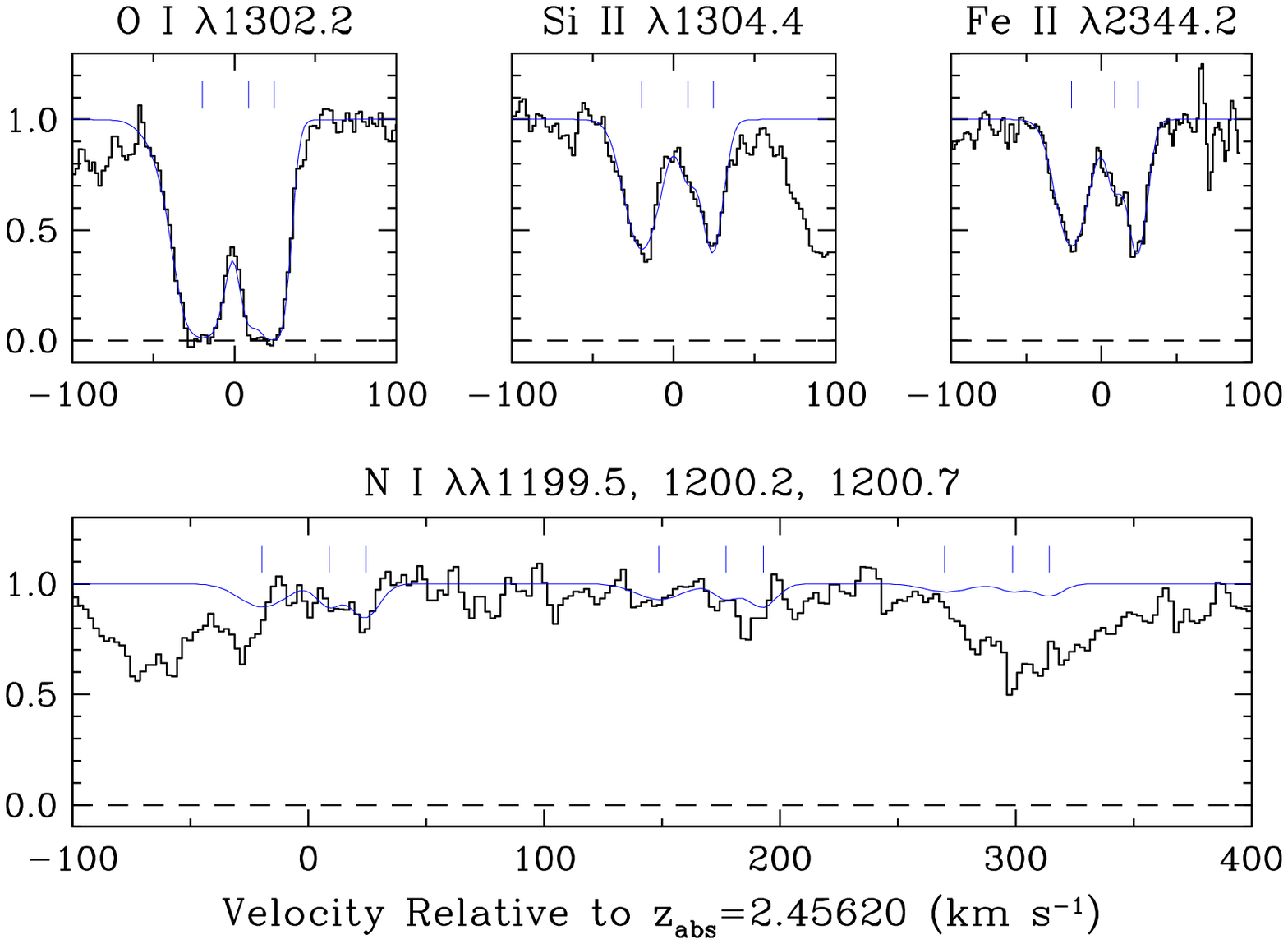}}}
\vspace{-8cm}
\caption{Normalised absorption profiles (histograms)
of selected metal lines in the $z_{\rm abs} = 2.45620$ DLA
in Q1409$+$095. The $y$-axis is residual intensity.
The thin continuous lines show theoretical profiles
produced by VPFIT with the parameters listed
in Table 2. Three velocity components, indicated
by vertical tick marks, contribute to the
absorption in this DLA.
}
\end{figure*}

\subsection{The $z_{\rm abs} = 2.66820$ sub-DLA in Q1409$+$095}

There is a second high column density 
system in Q1409$+$095; although
its $N$\/(H~I)$ = 5.0 \times 10^{19}$\,\cm\
is below the limit normally adopted for DLAs
(such systems are sometimes referred to as sub-DLAs),
the damping wings of the \lya\ line are easily recognisable
in the UVES spectrum (see Figure 1)
and $N$\/(H~I) is well determined.

The O~I, Si~II, and Fe~II absorption lines in this system
are well reproduced with a single absorption component
with a narrow velocity dispersion, $b = 7.5$\,\kms\
(see Figure 3); although there may be further substructure 
within this component, it is difficult to resolve it unambiguously
in our data. All three lines give consistent abundances of
between 1/15 and 1/20 of solar (see Table 3). Again, the 
N~I triplet is blended with \lya\ forest absorption. While
there are features which match the expected positions
of two of the N~I lines, we take the fit shown in Figure 3 as 
indicative of the maximum amount of N~I which could be present.
In this case, N is less abundant than 1/150 of solar.

% FIGURE 3

\begin{figure*}[h]
\centerline{\resizebox{18cm}{!}{\includegraphics{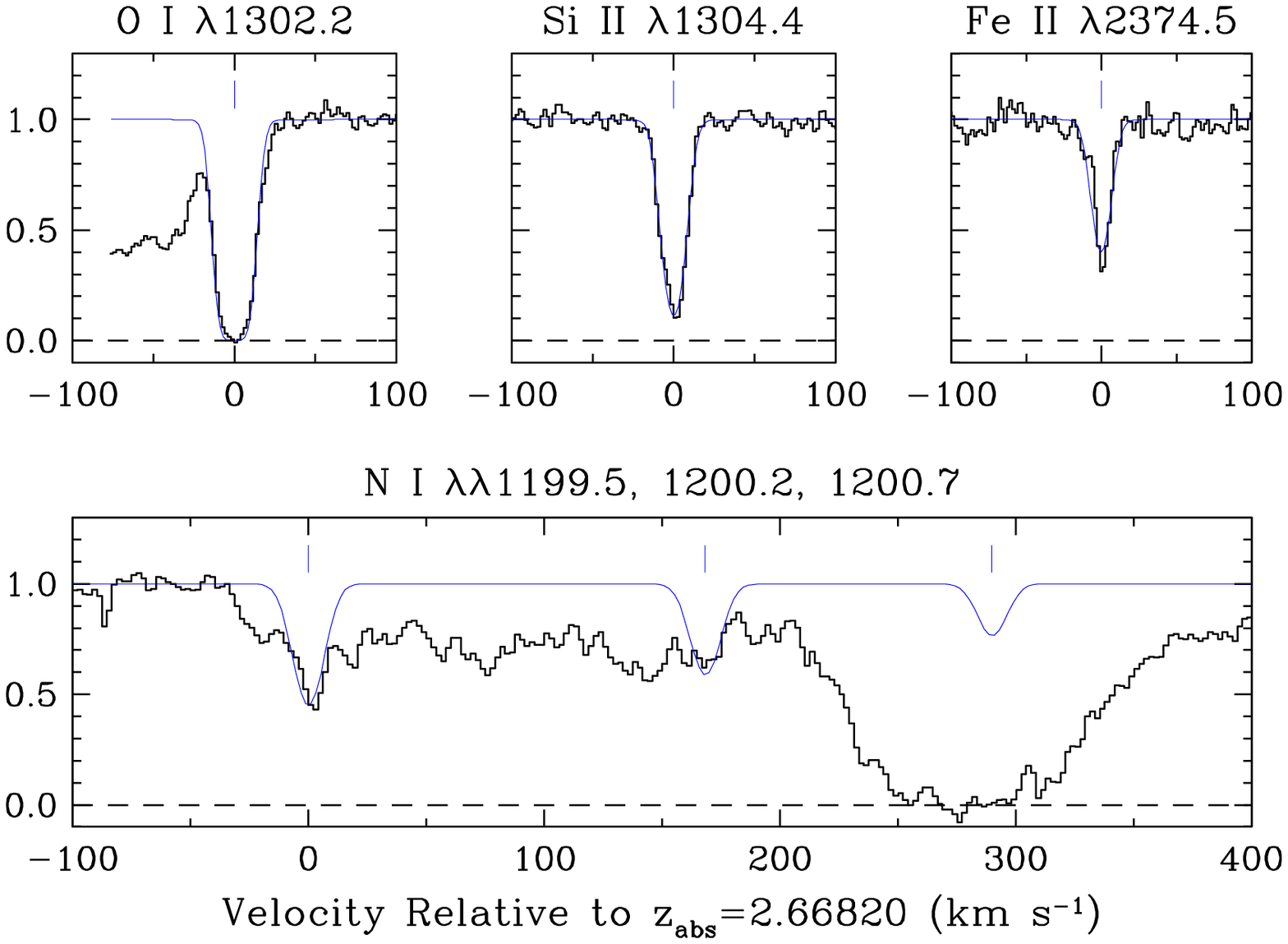}}}
\vspace{-8cm}
\caption{Normalised absorption profiles (histograms)
of selected metal lines in the $z_{\rm abs} = 2.66820$ sub-DLA
in Q1409$+$095. The $y$-axis is residual intensity.
The thin continuous lines show theoretical profiles
produced by VPFIT with the parameters listed
in Table 2. The fit to the N~I region 
(within the \lya\ forest) indicates 
the maximum amount of N~I which could be present.
}
\end{figure*}

\subsection{The $z_{\rm abs} = 2.08681$ DLA in Q1444$+$014}

Unlike the other DLAs in this study, this system exhibits a very complex
velocity structure. The fit to the Si~II~$\lambda 1304.4$ line requires at least
13 components spread over the velocity interval
$\Delta v \simeq 350$\,\kms, 
from $-170$ to $+176$\,\kms\ relative to $z_{\rm abs} = 2.08681$
(see Figure 4). Unfortunately, many of the components
are strongly saturated in O~I; while the weaker ones
are well matched by the model parameters returned by VPFIT for Si~II,
we cannot determine a reliable value of 
the total column density of O~I as it is dominated
by the heavily saturated absorption between $v \simeq 0$ and $-150$\,\kms.
In the bottom panel we show the partial fit, by the weaker components
only, to the region encompassing the N~I triplet. The combination
of 13 absorption components which partially overlap
in the closely separated triplet lines, together with blending
with \lya\ forest lines, makes it impossible to estimate the abundance of N.
As we are unable to determine (N/O) in this system, we do not consider it further
in this paper.

% FIGURE 4

\begin{figure*}[h]
\centerline{\resizebox{18cm}{!}{\includegraphics{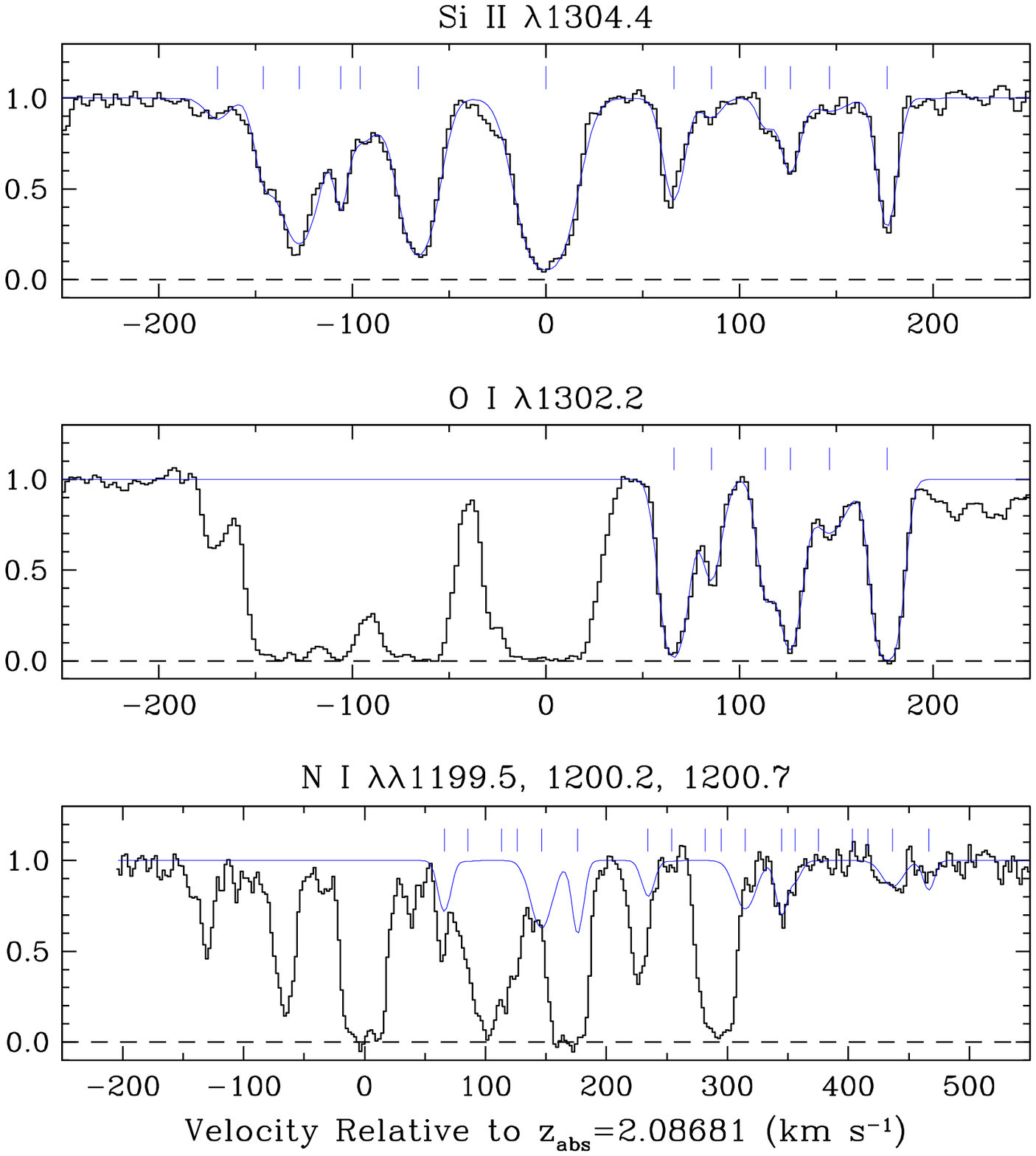}}}
\vspace{-3.5cm}
\caption{Normalised absorption profiles (histograms)
of selected metal lines in the $z_{\rm abs} = 2.08681$ DLA
in Q1444$+$014. The $y$-axis is residual intensity.
The velocity structure of this system is exceedingly
complex; at least 13 separate absorption components
(indicated by vertical tick marks in the top panel)
contribute to the absorption lines. While the profile
of Si~II~$\lambda 1304.4$ can be reproduced adequately
by VPFIT (thin continuous line), we cannot derive
reliable column densities for O~I nor N~I. In the former,
the strongest components are heavily saturated, while
the triplet structure of the latter leads to intractable
blending. The VPFIT profiles for O~I and N~I 
are partial fits to the weaker components only, shown 
for illustration.
}
\end{figure*}

\subsection{The $z_{\rm abs} = 2.07623$ DLA in Q2206$-$199}

This is a well-known DLA, one of the first to be studied
at echelle spectral resolution thanks to the 
brightness of the QSO (Pettini et al. 1990; Pettini \& Hunstead 1990).
It has a remarkably simple velocity structure---the absorption
apparently arises in a single component with $b \simeq 6$\,\kms\
(Figure 5 and Table 2).
This property makes it one of the
few DLAs suitable for measuring the abundance of deuterium
(Pettini \& Bowen 2001). It also exhibits a very low metallicity;
O, Si and Fe are below solar by factors of 90, 210, and 330 
respectively (Table 3).
For the elements in common (Si and Fe), our results
are in good agreement (within 0.1 dex) with 
the Keck data analysed by Prochaska \& Wolfe (1997).
As can be seen from Figure 5, we have a tentative detection
of N~I from which we deduce an upper limit
to the abundance of N of only 
$3 \times 10^{-4}$ of solar.

It is interesting that in this system we deduce
[O/Si]\,$ = +0.4 \pm 0.1$ (Table 3). Possibly, 
this is a hint of the trend of 
increasing [O/Si] with decreasing metallicity seen 
in Galactic halo stars with [Fe/H]\,$\simlt -2$
(Edvardsson et al. 1993; Ryan, Norris, \& Beers 1996;
Israelian et al. 2001).
Less interesting explanations are that Si is depleted onto
dust (although a depletion by a factor of 0.4 in the log
seems somewhat high, given the low metallicity
of this DLA---see Pettini et al. 1997) or that we have 
overestimated $N$(O~I) when fitting the saturated
$\lambda 1302$ absorption line.
The abundance of S, an undepleted element which tracks
O more closely than Si (Israelian \& Rebolo 2001; Takada-Hidai 
et al. 2001---see discussion in \S4) 
would help discriminate between these three
possibilities, but unfortunately  all three transitions of 
the S~II$\lambda1256$ multiplet in this system
are blended with \lya\ forest lines.

% FIGURE 5

\begin{figure*}[h]
\centerline{\resizebox{18cm}{!}{\includegraphics{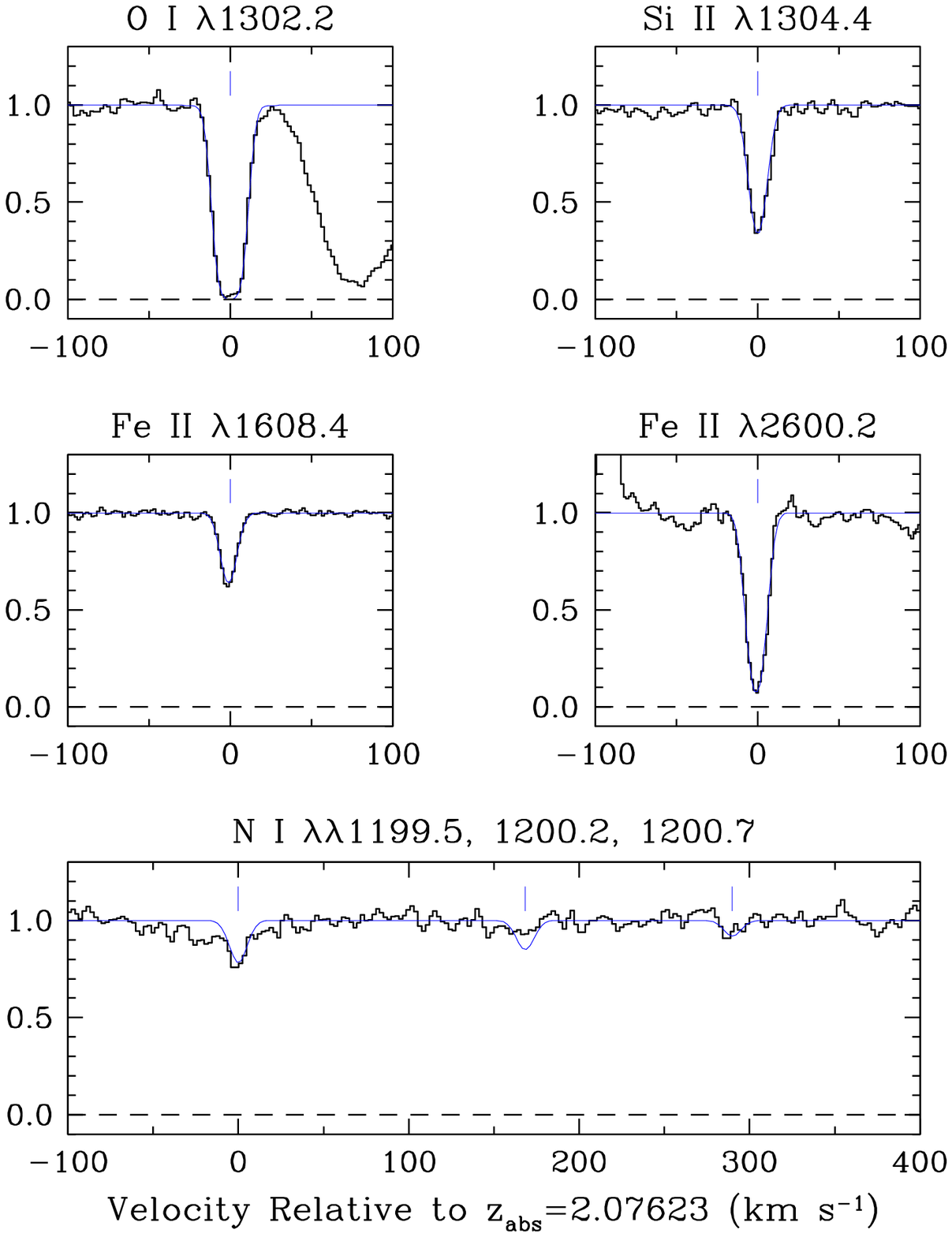}}}
\vspace{-4cm}
\caption{Normalised absorption profiles (histograms)
of selected metal lines in the $z_{\rm abs} = 2.07623$ DLA
in Q2206$-$199. The $y$-axis is residual intensity.
The thin continuous lines show theoretical profiles
produced by VPFIT with the parameters listed
in Table 2. The absorption seen in this DLA 
is due to a single velocity component with $b = 6.1$\,\kms.
}
\end{figure*}

\section{The DLA Sample}

In the last few years there have been several
new measurements of the abundance of N, generally 
reported as part of abundance studies of individual DLAs.
We have collected these data in Table 4 where references
to the original papers can also be found. 
All of the measurements in the table are from 
spectra of high signal-to-noise ratio and 
high (echelle) resolution obtained
with 8-10\,m class telescopes. For simplicity
we limit ourselves to absorption systems
with $N$(H~I)$ \geq 2 \times 10^{20}$\,cm$^{-2}$;
inclusion of similar data available for a few
sub-DLAs would not change any of our conclusions.

%
% TABLE 4
%___________________________________ Two column table (place early!)
\begin{table*}[!]
\begin{center}
\caption{N, O, S, and Fe Abundance Measurements in DLAs}
\begin{tabular}{llclccclcc}
\hline\hline
\noalign{\smallskip}
QSO 
& ~~~$z_{\rm abs}$ 
& log $N$\/(H~I)
& log $N$\/(N~I)
& log $N$\/(O~I)
& log $N$\/(S~II)
& (O/H)+12$^{a}$
& (N/O)$^{a}$ 
& (Fe/O)$^{a}$
& Ref.$^{b}$ \\

&
& (cm$^{-2}$)
& ~~~(cm$^{-2}$)
& (cm$^{-2}$)
& (cm$^{-2}$)
&
&
&
& 
\smallskip
\\
\hline
\\
Q0000$-$2620    & 3.3901   & 21.41 & 14.73           & 16.45    & \ldots   & 7.04 & $-1.72$ &  $-1.58$     &   1 \\
Q0100$+$1300    & 2.3090   & 21.32 & 15.29           & \ldots   & 15.12    & 7.34 & $-1.37$ &  $-1.56$     &   2, 3 \\
%HS0105$+$1619  & 2.53600  & 19.42 & 12.31           & 14.38    & \ldots   & 6.96 & $-2.07$ &  $-1.47$     &   3 \\
Q0201$+$1120    & 3.38639  & 21.26 & 15.33           & \ldots   & 15.21    & 7.49 & $-1.42$ &  $-1.40$     &   4 \\
J0307$-$4945    & 4.46658  & 20.67 & 13.57           & 15.91    & \ldots   & 7.24 & $-2.34$ &  $-1.70$     &   5 \\
Q0347$-$3819    & 3.02486  & 20.63 & 14.89           & 16.64    & \ldots   & 8.01 & $-1.75$ &  $-2.21$     &   6 \\
%Q0930$+$2858   & 3.2353   & 20.18 & 13.82           & \ldots   & 13.67    & 7.03 & $-1.39$ &  \ldots      &   2 \\
%Q1100$-$2629   & 1.83891  & 19.34 & $\leq 12.65^{c}$ & 14.64  & \ldots   & 7.30 & $\leq -1.55$ & \ldots  &   7 \\
Q1409$+$095     & 2.45620  & 20.54 & $\leq 13.19$    & 15.15    & \ldots   & 6.61 & $\leq -1.96$ & $-1.41$ &   7 \\
%Q1409$+$095    & 2.66820  & 19.70 & $\leq 13.51$    & 15.27    & \ldots   & 7.57 & $\leq -1.82$ & $-1.42$ &   7 \\
GB1759$+$7539   & 2.62528  & 20.76 & 14.99           & \ldots   & $15.21^c$    & 7.99 & $-1.76^c$ &  $-1.81$     &   8,9 \\
%GB1759$+$7539  & 2.91016  & 19.80 & 13.24           & \ldots   & 13.67    & 7.41 & $-1.97$ &  $-1.56$     &   8 \\
Q2206$-$199     & 2.07623  & 20.44 & $\leq12.88$     & 15.24    & \ldots   & 6.80 & $\leq -2.36$ &  $-1.87$     &   7 \\
Q2343$+$1230    & 2.4313   & 20.34 & 14.67           & \ldots   & 14.71    & 7.91 & $-1.58$ &  \dots       &   2 \\
Q2348$-$1444    & 2.27936  & 20.56 & $\leq 13.22$    & \ldots   & 13.73    & 6.71 & $\leq -2.05$ & $-1.48$ &   3 \\
\\
\hline
\end{tabular}
\begin{minipage}{160mm}
\smallskip
\item[$^{\rm a}$] When the oxygen abundance
is not available, S has been used as a proxy for O
by assuming the solar ratio (O/S)$_{\odot} = +1.54$ (Grevesse \&
Sauval 1998; Holweger 2001); see text (\S4) for justification.

\item[$^{\rm b}$] References---1: Molaro et al. (2000); 
2: Lu, Sargent, \& Barlow (1998);
3: Prochaska \& Wolfe (1999);
4: Ellison et al. (2001);
5: Dessauges-Zavadsky et al. (2001);
6: Levshakov et al. (2002);
7: This work;
8: Outram, Chaffee, \& Carswell (1999);
9: Prochaska et al. (2002).

\item[$^{\rm c}$] Prochaska et al. (2002) found that only about half of the metals
in this DLA arise in a neutral component, while the reminder are in partially
ionised gas. It is unclear what fraction of S in the ionised gas is S~II;
in the extreme case that half of the observed S~II is in an ionised region
where O~I and N~I are absent, the value of (N/O) for this DLA should be 
increased by +0.3\,dex.

%\item[$^{\rm c}$] Main absorption component only.

\end{minipage}
\end{center}
\end{table*}

% ------------------------------------------------------------------

The full sample consists of 10 DLAs 
in which the abundance of N has been measured (seven detections and 
three upper limits). In one half of the cases the abundance of O is
available directly, either because the O~I~$\lambda 1302.2$ line
is not strongly saturated, or from weaker, unsaturated, transitions 
to higher energy levels. In the other five cases we take $N$\/(S~II)
as a proxy for $N$\/(O~I), after correcting for the solar abundance ratio
(O/S)$_{\odot} = +1.54$ (Grevesse \& Sauval 1998; Holweger 2001).
The justification for this approach is as follows.

First, in Galactic stars the (O/S) ratio
appears to be approximately solar
over a wide range of metallicities down to
[O/H]\,$ \simlt -2$ (Israelian et al. 2001; Israelian \& Rebolo 2001;
Takada-Hidai et al. 2001;
Nissen et al. in preparation).
Second, in the diffuse Galactic interstellar medium both elements 
show little affinity for dust (Savage \& Sembach 1996);
dust corrections for O and S are 
almost certainly unimportant for our purposes
at the reduced levels of dust depletions typical of metal-poor
DLAs (Pettini et al. 1997; Vladilo 1998).
Finally, differential ionisation corrections between O~I and S~II
are likely to be small. Theoretically, the photoionisation
calculations by Vladilo et al. (2001) show that, for all
the models they considered,
if log~(O/S)\,=\,log~[$N$\/(O~I)/$N$\/(S~II)$]+{\cal C}$,
then $|{\cal C}| \simlt 0.2$ ($\simlt 0.1$) 
when log~$N$\/(H~I)$\geq  20.2$ ($\geq 20.6$).
This condition is met by 5 (4) out of the 5 DLAs in Table 4
for which we use S as a proxy for O. 
%In the seventh
%case, the sub-DLA at $z_{\rm abs} = 2.91016$ in GB1759$+$7539,
%we may have overestimated the abundance of O by using S.
For all the DLAs considered here $|{\cal C}|$[N/O]\,$\leq 0.002$
so that log~(N/O)\,=\,log~[$N$\/(N~I)/$N$\/(O~I)].

Some authors (e.g. Lu et al. 1998) have also used Si
to deduce the abundance of O, by assuming that in DLAs 
$N$(O~I)/$N$(Si~II)\,=\,(O/Si)$_{\odot}$.
This assumption, however, is less secure
than the equivalent one for S and has
been criticised, for example, by Matteucci,
Molaro, \& Vladilo (1997). 
First, Si can be depleted onto dust.
Second, it is unclear whether the abundances
of O and Si do track each other at low
metallicities (Edvardsson et al. 1993; 
Ryan, Norris, \& Beers 1996;
Israelian et al. 2001); theoretically
they are not expected to, if some of the 
Si is produced by type Ia supernovae
(Matteucci et al. 1997).
Neither of these complications applies to 
S. In the past, the use of Si as an indicator
of the O abundance had been motivated by
the paucity of measurements of [O/H] and 
[S/H] in DLAs; with the larger body of data
now available this approach is no longer necessary.
For this reason, we restrict ourselves to
O and S in the following analysis.

\subsection{Nitrogen and Oxygen in DLAs}

We are now ready to compare the abundances of N and O in 
high redshift DLAs with those measured in H~II regions
in the local universe (Figure 6). For the latter, we use the
compilation assembled by Henry et al. (2000---references to 
the original surveys are given in the figure caption), augmented
by some additional work (Ferguson, Gallagher, \& Wyse 1998;
van Zee 2000).
The dashed lines in the figure are approximate boundaries
of the region in the (N/O) vs. (O/H) plot where we may expect
DLAs to fall. The line labelled `Secondary' is simply
an extrapolation to low (O/H) values
of the (N/O) vs. (O/H) trend at high metallicities,
where secondary N presumably dominates. 
{\it If} the extrapolation to low metallicities
is valid (we do not know that it is), this is the minimum
amount of N (relative to O) which one may expect to find,
irrespectively of the previous star formation history of
the system under consideration. The `Primary' level
is just the plateau at (N/O)\,$ = -1.5$ indicated by existing
measurements in low metallicity H~II regions.
We stress that these are just approximate empirical limits
which are not derived from a full chemical evolution treatment
of the available data (e.g. Larsen, Sommer-Larsen, \& Pagel 
2001)

As can be seen from Figure 6, the DLAs in our sample
do indeed populate the region of the diagram 
delimited by the broken lines. Six of the DLAs 
considered exhibit levels of N enrichment which are
similar to those of the most metal-poor H~II regions.
On the other hand, in four cases N appears to be 
below the `Primary' line, by up to nearly one order of
magnitude.

% FIGURE 6

\begin{figure*}  %* to go across two columns
%\vspace{-3.0cm}
\centerline{\includegraphics[width=40pc,angle=270]{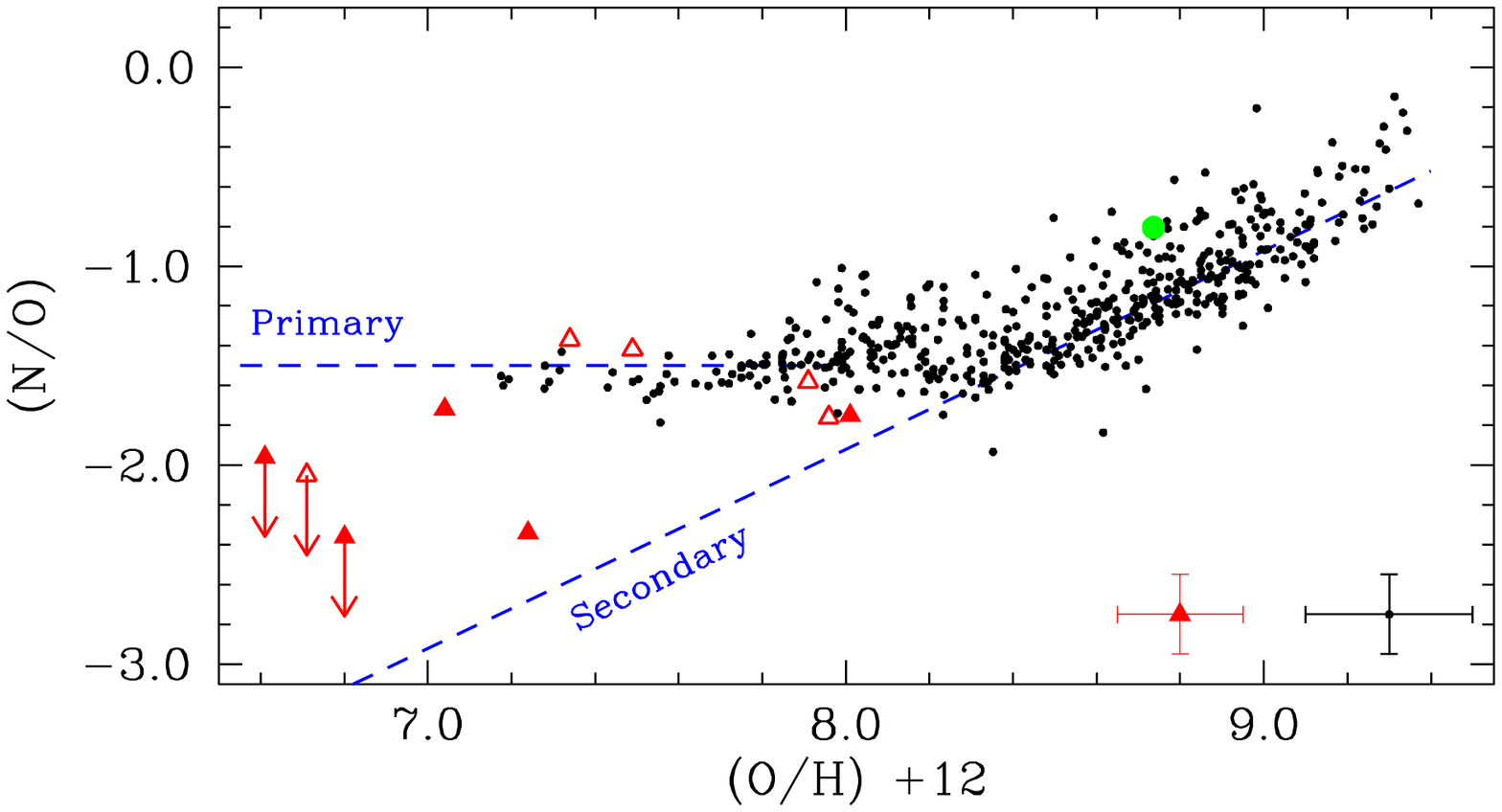}}
\vspace{-4cm}
\caption{Abundances of N and O in extragalactic H~II regions
(small dots) and damped \lya\ systems (large triangles).
The H~II region measurements are from the studies by 
Kobulnicky \& Skillman (1996); 
Thurston, Edmunds, \& Henry (1996);
van Zee et al. (1998); Ferguson, Gallagher, \& Wyse (1998);
Izotov \& Thuan (1999); 
and van Zee (2000).
The values for damped \lya\ systems
are from the compilation in Table 4; filled 
triangles denote DLAs where the abundance of O 
could be measured directly, while open triangles
are cases where S was used as a proxy for O (see text).
%Sub-DLAs are shown as inverted triangles.
The error bars in the bottom right-hand corner
give an indication of the typical uncertainties;
the large dot corresponds to the 
solar abundances of N and O from the recent reappraisal
by Holweger (2001).
The dashed lines are approximate
representations of the primary and secondary 
levels of N production (see text).
}
\end{figure*}

\section{Discussion}

The results presented here, in Figure 6 and Table 4,
confirm the earlier conclusion by Lu et al. (1998)
that the (N/O) ratio in DLAs exhibits a larger
range of values than that of any other pair of heavy 
elements which are not depleted onto dust
(Prochaska \& Wolfe 2002).
We now consider possible causes of this scatter.

The simplest explanation may be just measurement
error. Although the errors on 
$N$\/(N~I) and $N$\/(O~I) quoted by the sources
referenced in Table 4 are
normally less than 0.1\,dex (and always less 
than 0.2\,dex), it must be remembered
that the N~I lines (a) are generally weaker than most
other metal transitions used for abundance work in DLAs,
and (b) occur within the \lya\ forest
where blending can complicate their analysis.
It is thus conceivable that the true uncertainties,
particularly in the N abundance, may have been
underestimated. On the other hand, if measurement
error were the sole cause of the dispersion 
of the DLA points in Figure 6, it would be a remarkable
coincidence that all 10 values in our sample
should fall within the boundaries delimited
by the Primary and Secondary lines in the figure.
In principle one would expect measurement errors
to scatter points above the Primary line and,
at the larger values of (O/H), below the
Secondary boundary. For many of the low
points in Figure 6, it would be very hard
to understand how an absorption line
$\sim 3-5$ times stronger than observed could have been missed.
We conclude, therefore,
that measurement errors are not the principle cause
of the dispersion observed.

We do not subscribe to the view proposed
by Izotov et al. (2001) that unrecognised ionisation
corrections are responsible for the low (N/O)
values. As explained above,
in about half of the cases (N/O)$_{\rm DLA}$
is comparable to the value measured in nearby metal-poor
H~II regions; there is no evidence that these DLAs
are different from the rest of the sample in their
level of ionisation.
Similarly, there is no separation
in Figure 6 between DLAs where (O/H) has been
measured directly and those where it has been deduced
by reference to S.

Some may be concerned about possible systematic differences 
between abundances measured via absorption lines in
cool interstellar H~I clouds and those deduced from 
the analysis of nebular emission lines from H~II gas.
However, there are no precedents for such systematic
differences in the solar vicinity 
(Meyer, Cardelli, \& Sofia 1997; Meyer, Jura, \& Cardelli 1998)
and, in any case, here we are dealing with a scatter
rather than a systematic offset.

Qualitatively, our findings are consistent with,
and indeed provide empirical evidence in favour of,
an origin of primary nitrogen in intermediate mass stars.
The observed range in the (N/O) ratio at low (O/H)
is, in this picture, a natural consequence of the
delayed release of N into the ISM relative to the
O produced by Type II supernovae. However, there
are puzzling aspects in this scenario too, in that
the proportion of DLAs with values of (N/O)
below the primary level seems at first sight
to be surprisingly high.
Recall that Henry et al. (2000) estimated the 
time delay between N and O release into the ISM
to be $\sim 250$\,Myr, if the main
source are stars in the mass range $7 - 4\,M_{\odot}$.
The median redshift
of the DLA sample in Table 4 is $\langle z \rangle = 2.5$;
in a $\Omega_{\rm M} = 0.3$, $\Omega_{\Lambda} = 0.7$,
$h = 0.65$ cosmology this corresponds to a 
look-back time of $11.7$\,Gyr. \footnote{The age
of the universe in this cosmology is $14.5$\,Gyr.}
There is now evidence
that star formation in the universe was already at high
levels well before this epoch, certainly 
by $z \simgt 4.5$ (Steidel et al. 1999),
and possibly up to $z \simeq 6$ (Shapley et al. 2001;
Becker et al. 2001; Dawson et al. 2001).
In this cosmology, $z = 6$ corresponds to
a look-back time of $\sim 13.5$\,Gyr.
Thus, the time interval 
between $z = 6$ and our median $\langle z_{\rm DLA}\rangle  = 2.5$
is seven times higher than the estimated time delay for the
release of primary nitrogen.
If the galaxies giving rise to DLAs formed
continuously between $z = 6$ and 2.5, naively one may 
expect to find only one DLA out of seven in the
interim period when primary nitrogen has not yet
caught up with the oxygen released by the same
episode of star formation. Instead we find at least
four such cases out of ten (see Figure 6).

Perhaps we are just seeing
the effects of small number statistics
and a larger sample of (N/O) measurements
will in future reveal that the present 40\% 
is an overestimate of the number of DLAs 
with a N deficiency.
However, if such a high fraction is confirmed by
future observations, we can think of two explanations
for this apparent puzzle. The four DLAs with values 
of (N/O) clearly below
the primary level in Figure 6 have abundances
(O/H) less than  $\sim 1/30$ of solar.
Thus, they have presumably experienced little star formation 
up to the time when we observe them;
in a closed box model, they would have turned less than
1/30 of their gas into stars. It therefore seems at least
plausible that these DLAs arise preferentially in 
young galaxies, which have only recently condensed
out of the intergalactic medium. In other words,
the very fact that the difference between
primary and secondary nitrogen is most pronounced at
low metallicities introduces a bias in our sample
such that it may include a relatively high proportion
of galaxies caught in the initial stages of their
chemical evolution (e.g. Cen et al. 2002).

A second possibility is that
the time delay for the release of primary
nitrogen may be longer when metal abundances are lower.
This dependence may arise if at lower metallicities
stars of progressively lower masses can synthesize 
and release nitrogen,
because their interiors are hotter
(Lattanzio \& Forestini 1999) and/or
through the increasing importance of rotation 
(Meynet \& Maeder 2002a,b; see also Marigo 2001). 
The slope of the initial mass function would
then shift the peak of the N production
to lower mass stars, with longer evolutionary times.
While these effects remain to 
be fully quantified,
there have been suggestions that at the metallicities
of DLAs the full release of primary N
may lag behind that of O by up to 800\,Myr
(Lattanzio et al., in preparation).
If confirmed, this proposal would go a long way towards
explaining the high proportion of DLAs 
which are underabundant in N at $\langle z \rangle = 2.5$.

\subsection{Clues from the Abundance of Iron?}

It should be possible to test the hypothesis
that the dispersion of (N/O) values in DLAs
is a consequence of the 
delayed release of primary nitrogen by considering the 
abundances of other elements synthesized by low mass stars.
Observationally, iron is one of the more easily accessible 
through several rest frame ultraviolet transitions. Measurements of 
(Fe/H) are available for
9 out of the 10 DLAs in our sample\footnote{The 
same is not true for other Fe-peak elements,
especially Zn whose abundance may be easier 
to interpret because, unlike Fe, Zn is not depleted onto dust.};
the corresponding values of (Fe/O) are listed
in Table 4.

It is generally thought that approximately 2/3
of the iron is produced in Type Ia supernovae.
While it is still unclear which types of binary system
give rise to such events, there seems to be a consensus
that the evolutionary timescales of Type Ia
SN progenitors are of the same order as, or longer than,
those of the $7-4\,M_{\odot}$ stars which are the source
of primary nitrogen (e.g. Matteucci \& Recchi 2001).
From this it follows that DLAs deficient in N 
(in the sense of having less N than the primary level)
should also exhibit sub-solar (Fe/O) ratios.
We examine this point in Figure 7, by plotting
the quantity [(N/O) + 1.5] (that is, the difference
between the measured (N/O) and the primary
plateau at (N/O)\,$ = -1.5$) vs. 
[(Fe/O) + 1.29] (the difference
between observed and solar (Fe/O)
adopting (Fe/O)$_{\odot} = -1.29$ from Holweger 2001).

% FIGURE 7

\begin{figure*}[h]
\vspace{-2cm}
\centerline{\resizebox{18cm}{!}{\includegraphics{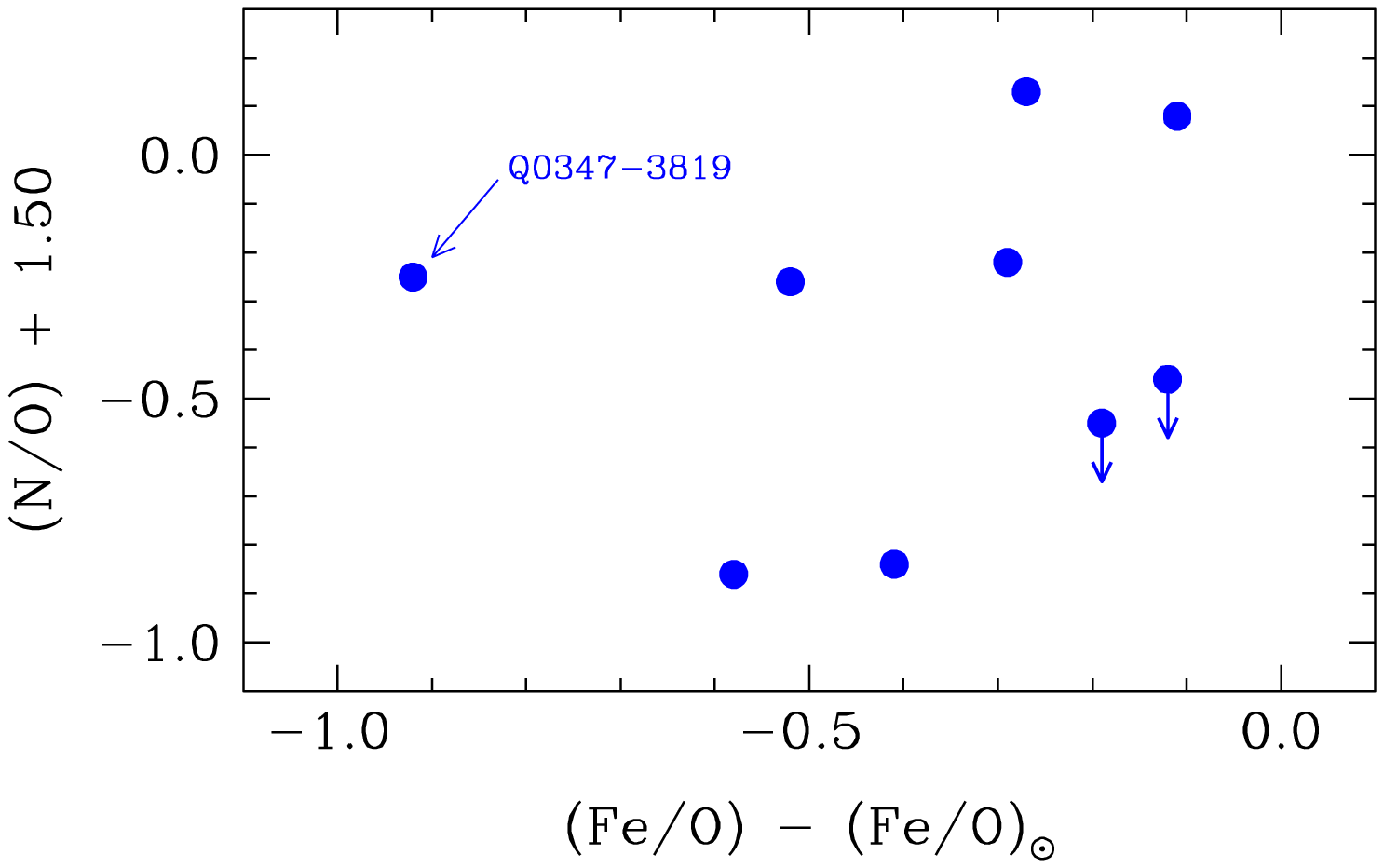}}}
\vspace{-6cm}
\caption{In this figure we plot the deficiency of N relative to
the Primary level at (N/O)\,$ = -1.5$
vs. [Fe/O], the departure of the (Fe/O) ratio in DLAs
from the solar (Fe/O)$_{\odot} = -1.29$ (Holweger 2001).
In the logarithmic units used here, the abundance ratio
of two elements has a typical measurement error of between
0.1 and 0.2\,. Measured values of (N/O) and (Fe/O) 
for the nine DLAs shown are listed in 
columns (8) and (9) of Table 4.
}
\end{figure*}

There are two complications which affect the interpretation
of the data in Figure 7. First, if massive stars 
which explode as Type II SN also produce some 
Fe, the effect of time delay will be less 
pronounced for Fe than for N.
Indeed, Galactic stars
with metallicities similar to those of the DLAs
considered here, between $\sim 1/10$ and $\sim 1/100$ of solar, 
are deficient in Fe only by factors of 2--3.
In the units of Figure 7, 
[Fe/O]\,$\equiv$\,(Fe/O)\,$-$\,(Fe/O)$_{\odot} = -0.3$ to $-0.5$
(e.g. Israelian et al. 2001). This is larger, but not by much, than
the typical accuracy of 0.1--0.2 dex with which this ratio is
measured in DLAs.
A more serious problem is that, unlike O, Fe can be depleted onto
dust and this effect will mimic the 
intrinsic underabundance which we are 
testing for. The combination of the subtlety of the effect
with the uncertainty in correcting for dust depletions may well
explain, at least in part, why the $\alpha$-element 
enhancement well established
in Galactic metal-poor stars has proved so difficult to pin down in DLAs
(Vladilo 1998; Pettini et al. 2000; Prochaska \& Wolfe 2002, 
Ledoux, Bergeron, \& Petitjean 2002).
Accounting for the fraction of Fe in solid form
requires taking into consideration the full complement
of abundance measures in a DLA (e.g. Vladilo 2002), 
and is beyond the scope of this work.
The values of [Fe/O] plotted in Figure 7 
should therefore be considered as indicative
of the maximum intrinsic deficiency of Fe relative to O
in the DLAs in our sample.

These complications make it difficult to draw firm
conclusions from the results in Figure 7. 
Some of the DLAs appear to meet our expectations,
by exhibiting low values of (N/O) matched by corresponding 
Fe underabundances. Similarly, DLAs which lie closer
to the Primary line at (N/O)\,$= -1.5$ 
show a range of [Fe/O] values, as expected 
if the time delay for the release of Fe is longer than
that of N. But there are also cases which evidently
do not fit this pattern. The DLA in Q0347$-$3819
may be a special case, in that it is the 
{\it only} DLA known to date with such a strong
$\alpha$-element enhancement (Levshakov et al. 2002).
More problematic, however, are the two
DLAs with low (N/O) (both are upper limits)
and only a minor (if any) Fe deficiency.
Possibly we have underestimated the N abundance in
these cases because of blending with \lya\ forest
lines.\footnote{A similar apparent inconsistency 
was pointed out by Centuri\'{o}n et al.
(1998) for the DLA in Q0100$+$1300.
The HIRES observations
of this QSO by Lu et al. (1998) 
and Prochaska \& Wolfe (1999),
used here, give (N/O)\,+\,1.5$ = +0.13$ and
[Fe/O]$\,=-0.27$, which are consistent with 
our working hypothesis. Centuri\'{o}n et al. (1998),
on the other hand, derived a much lower column 
density of N---by a factor of four---from observations
of the weaker N~I~$\lambda\lambda1134$ triplet
with the 4\,m William Herschel telescope on la Palma.
Observations of the N~I~$\lambda\lambda1134$ triplet
with an 8-10\,m telescope are required to resolve
this discrepancy.
}

We conclude that the results of Figure 7 do not provide
clear support for the scenario where both N and Fe
lag behind the release of O into the ISM. Possibly,
as more data become available, a trend between 
(N/O) and (Fe/O) will emerge, if one exists.
But Figure 7 may also be telling us that
there is a limit to the degree to which
our ideas on the chemical evolution of the Milky Way
can be applied to the high redshift DLA galaxies.

\section{Summary and Conclusions}

We have presented new measurements of the abundances of
N, O, Si, and Fe in two metal-poor damped \lya\ systems
(plus one sub-DLA) obtained with UVES on the VLT. 
We have combined these
data with others from the literature to form a sample
of 10 DLAs for which reliable measurements of the 
abundance of N (or useful upper limits) are available.
The sample consists exclusively of high resolution, high
S/N measurements obtained
with 8-10\,m class telescopes. Our aim was to test
current ideas on the nucleosynthesis of N, by
extending measurements of the abundances of N and O
from H~II regions in the local universe 
to the younger and less enriched galaxies 
at high redshift ($\langle z \rangle = 2.5$)
that give rise to the DLAs.
Our principal conclusions are as follows.

\begin{enumerate}

\item The (N/O) ratio remains a difficult quantity to measure,
because the absorption lines available generally have widely
different optical depths, so that when the N~I~$\lambda 1200$
triplet is detected O~I~$\lambda 1302.2$ is saturated.
Furthermore, the N~I lines are often blended within
the \lya\ forest. We have not identified an optimum strategy
for dealing with these difficulties. DLAs at higher
redshifts than those shown here make accessible
weaker transitions of O~I at far-UV wavelengths, but
the chances of confusion of N~I with \lya\ forest
lines are then increased.

\item We have argued that S, when available, can be used
as a proxy for O, because they are both undepleted
$\alpha$-elements and the differential ionisation 
corrections between O~I and S~II should not
be important when the column density of neutral hydrogen
is as high as in DLAs.
However, it would obviously be reassuring to confirm
empirically that (O/S)\,$\equiv N$\/(O~I)/$N$\/(S~II)
in at least a few cases,
with specially targeted observations.

\item We confirm earlier reports that the (N/O) ratio
shows a larger dispersion of values than other
ratios of heavy elements in DLAs. All 10 measurements
in our sample fall within the region in the
(N/O) vs. (O/H) plot bounded by the primary
and secondary levels of N production.

\item We have considered several possible explanations
for this scatter and conclude that the most plausible one
is that we are seeing the effects of a time delay,
by about 250\,Myr, 
in the release of primary N from $7 - 4\,M_{\odot}$
stars relative to the massive stars which release O
when they explode as Type II supernovae. 
Thus, our results provide empirical evidence in support of 
currently favoured ideas for the nucleosynthesis of N.
The uniform value (N/O)\,$\simeq -1.5$ (on a log scale)
seen in nearby metal-poor star-forming galaxies
can be understood in this scenario if these
galaxies are not young,
but contain older stellar populations,
as indicated by a number of imaging studies
with {\it HST}.

\item A surprisingly high proportion (40\%) 
of DLAs in our sample have apparently not
yet attained the full primary level of N enrichment
at (N/O)\,$\simeq -1.5$\,. 
%In retrospect, this is perhaps not surprising. 
Possibly,  the low metallicity regime---where
the difference between secondary and primary nitrogen enrichment
is most pronounced---preferentially selects
young galaxies which have only recently
condensed out of the intergalactic medium and begun forming
stars. A more speculative alternative, which needs to
be explored computationally, is that at low
metallicities stars with masses lower than 
$4 M_{\odot}$ may make a significant contribution
to the overall N yield.
The release of primary N may, under these circumstances,
continue for longer than 250\,Myr, perhaps for a substantial
fraction of the Hubble time at the median $\langle z \rangle = 2.5$
of our sample.

\item In this scenario, the abundance of Fe---two 
thirds of which is presumed to originate from Type Ia 
supernovae---should also lag behind that of O, 
by at least as long as that of N. 
However, conflicting clues are provided
by a plot (N/O) vs. (Fe/O) for the present data; 
while some DLAs do match the predictions of our working
hypothesis, there also appear to be some deviant cases.
The interpretation is complicated by the fact that
some fraction of the Fe is
also produced by massive stars and that
depletion onto dust can be an issue. 
A larger sample of measurements is required to 
assess whether the relative abundances of N, O and Fe
in DLAs are consistent with current ideas of chemical evolution
in galaxies.

\item Finally, we point out that the large dispersion in the abundance of 
N at these low metallicities essentially precludes the use of N~V
$\lambda\lambda 1238, 1242$ doublet in constraining 
models of the ionisation of the intergalactic medium
from the ratios of different ions in the \lya\ forest
(e.g. Boksenberg et al., in preparation).
\end{enumerate}

\begin{acknowledgements} 

It is a pleasure to acknowledge the 
competent assistance with the observations
by the ESO support staff at Paranal.
We are very grateful to Dick Henry for kindly providing us with
electronic tables of most of the data 
(on local H~II region abundances) plotted in Figure 6, and
to Jason X. Prochaska for making
widely available his comprehensive compilation
of DLA measurements and related atomic parameters.
Special thanks are due to David Bowen for running his
Monte Carlo simulations on our behalf.
Mike Edmunds, Paolo Molaro, Bernard Pagel and Sam Rix made valuable
comments on early versions of the manuscript.
This work was supported in part by the European RTN programme
``The Intergalactic Medium''.
Max Pettini thanks the Instituto de Astrofisica de Canarias 
and the Instituto de Astronom\'{\i}a, UNAM for their
hospitality during visits when this work was completed.

\end{acknowledgements}

\end{document}